\documentstyle [12pt] {article}

\catcode`@=12
\begin{document}
\thispagestyle{empty}
\begin{flushright} UCRHEP-T158\\March 1996\
\end{flushright}
\vspace{0.5in}
\begin{center}
{\Large	\bf Efficacious Extra U(1) Factor for\\}
\vspace{0.1in}
{\Large \bf the Supersymmetric Standard Model\\}
\vspace{1.5in}
{\bf E. Keith and Ernest Ma\\}
\vspace{0.3in}
{\sl Department of Physics\\}
{\sl University of California\\}
{\sl Riverside, California 92521\\}
\vspace{1.5in}
\end{center}

\begin{abstract}\
The totality of neutrino-oscillation phenomena appears to require the 
existence of a light singlet neutrino.  As pointed out recently, this 
can be naturally accommodated with a specific extra U(1) factor 
contained in the superstring-inspired E$_6$ model and its implied particle 
spectrum.  We analyze this model for other possible consequences.  
We discuss specifically the oblique corrections from Z-Z' mixing, the 
phenomenology of the two-Higgs-doublet sector and the associated neutralino 
sector, as well as possible scenarios of gauge-coupling unification.
\end{abstract}

\newpage
\baselineskip 24pt
\section{Introduction}

There are experimental indications at present for three types of neutrino 
oscillations: solar\cite{1}, atmospheric\cite{2}, and laboratory\cite{3}. 
Each may be explained in terms of two neutrinos differing in the square of 
their masses by roughly $10^{-5}$ eV$^2$, $10^{-2}$ eV$^2$, and 1 eV$^2$ 
respectively.  To accommodate all three possibilities, it is clear that 
three neutrinos are not enough.  On the other hand, the invisible width 
of the $Z$ boson is saturated already with the three known neutrinos, each 
transforming as part of a left-handed doublet under the standard electroweak 
$SU(2) \times U(1)$ gauge group.  There is thus no alternative but to assume 
a light singlet neutrino which also mixes with the known three doublet 
neutrinos.  As pointed out recently\cite{4}, this can be 
realized naturally with a specific extra U(1) factor contained in the 
superstring-inspired E$_6$ model and its implied particle spectrum.

In Section 2 we map out the essential features of this supersymmetric 
$SU(3)_C \times SU(2)_L \times U(1)_Y \times U(1)_N$ model.  In Section 3 
we study the mixing of the standard $Z$ boson with the $Z'$ boson required 
by the extra $U(1)_N$.  We derive the effective contributions of this mixing 
to the electroweak oblique parameters $\epsilon_{1,2,3}$ or $S,T,U$, and show 
that the $U(1)_N$ mass scale could be a few TeV.  In Section 4 we discuss 
the reduced Higgs potential at the electroweak scale and show how the 
two-Higgs-doublet structure of this model differs from that of the minimal 
supersymmetric standard model (MSSM).  In Section 5 we consider the 
neutralino sector and show how the lightest supersymmetric particle (LSP) 
of this model is constrained by the Higgs sector.  In Section 6 we venture 
into the realm of gauge-coupling unification and propose two possible 
scenarios, each with some additional particles.  Finally in Section 7 there 
are some concluding remarks.

\section{Description of the Model}

The supersymmetric particle content of this model is given by the fundamental 
{\bf 27} representation of E$_6$.  Under $SU(3)_C \times SU(2)_L \times U(1)_Y 
\times U(1)_N$, the individual left-handed fermion components transform as 
follows\cite{4}.
\begin{equation}
(u,d) \sim (3;2,{1 \over 6};1), ~~~ u^c \sim (3^*;1,-{2 \over 3};1), ~~~ 
d^c \sim (3^*;1,{1 \over 3};2),
\end{equation}
\begin{equation}
(\nu_e,e) \sim (1;2,-{1 \over 2};2), ~~~ e^c \sim (1;1,1;1), ~~~ N \sim 
(1;1,0;0),
\end{equation}
\begin{equation}
(\nu_E,E) \sim (1;2,-{1 \over 2};-3), ~~~ (E^c,N_E^c) \sim (1;2,{1 \over 2};
-2),
\end{equation}
\begin{equation}
h \sim (3;1,-{1 \over 3};-2), ~~~ h^c \sim (3^*;1,{1 \over 3};-3), ~~~ 
S \sim (1;1,0;5).
\end{equation}
As it stands, the allowed cubic terms of the superpotential are 
$u^c (u N_E^c - d E^c)$, $d^c (u E - d \nu_E)$, $e^c (\nu_e E - e \nu_E)$, 
$S h h^c$, $S (E E^c - \nu_E N_E^c)$, and $N (\nu_e N_E^c - e E^c)$, as well 
as $h u^c e^c$, $h d^c N$, and $N^3$.  We now impose a $Z_2$ discrete symmetry 
where all superfields are odd, except \underline {one} copy each of 
$(\nu_E, E)$, 
$(E^c, N_E^c)$, and $S$, which are even.  This gets rid of the cubic terms 
$h u^c e^c$, $h d^c N$, and $N^3$, but allows the quadratic terms $h d^c$, 
$\nu_e N_E^c - e E^c$, and $N^2$.

The bosonic components of the even superfields serve as Higgs bosons which 
break the gauge symmetry spontaneously.  Specifically, $\langle \tilde S 
\rangle$ breaks $U(1)_N$ and generates $m_h$ and $m_E$; the electroweak 
$SU(2)_L \times U(1)_Y$ is then broken by two Higgs doublets as in the 
MSSM, with $\langle \tilde N_E^c \rangle$ responsible for $m_u$, $m_D$, and 
$m_1$, and $\langle \tilde \nu_E \rangle$ for $m_d$, $m_e$, and $m_2$.  The 
mass matrix spanning the fermionic components of $\nu_e$, $N$, and the 
\underline {odd} $\nu_E$, $N_E^c$, and $S$ is then given by
\begin{equation}
{\cal M} = \left[ \begin{array} {c@{\quad}c@{\quad}c@{\quad}c@{\quad}c} 
0 & m_D & 0 & m_3 & 0 \\ m_D & m_N & 0 & 0 & 0 \\ 0 & 0 & 0 & m_E & m_1 \\ 
m_3 & 0 & m_E & 0 & m_2 \\ 0 & 0 & m_1 & m_2 & 0 \end{array} \right],
\end{equation}
where the mass term $m_N$ is expected to be large because $N$ is 
trivial under $U(1)_N$ and may thus acquire a large Majorana mass through 
gravitationally induced nonrenormalizable interactions\cite{5}, and 
$m_3$ comes from the allowed quadratic term $\nu_e N_E^c - e E^c$. 
This means that the usual seesaw mechanism holds for the three doublet 
neutrinos: $m_\nu \sim m_D^2/m_N$, whereas the two singlet neutrinos have 
masses $m_S \sim 2 m_1 m_2/m_E$ and mix with the former through $m_3$.
Note that ${\cal M}$ is really a $12 \times 12$ matrix because there are 
3 copies of ($\nu_e, N$) and 2 copies of ($\nu_E, N_E^c, S$).

\section{Z-Z' Mixing}

Let the bosonic components of the even superfields $(\nu_E, E)$, 
$(E^c, N_E^c)$, and $S$ be denoted as follows:
\begin{equation}
\tilde \Phi_1 \equiv \left( \begin{array} {c} \bar \phi_1^0 \\ - \phi_1^- 
\end{array} \right) \equiv \left( \begin{array} {c} \tilde \nu_E \\ \tilde E 
\end{array} \right), ~~~ \Phi_2 \equiv \left( \begin{array} {c} \phi_2^+ \\ 
\phi_2^0 \end{array} \right) \equiv \left( \begin{array} {c} \tilde E^c \\ 
\tilde N_E^c \end{array} \right), ~~~ \chi \equiv \tilde S.
\end{equation}
The part of the Lagrangian containing the interaction of the above Higgs 
bosons with the vector gauge bosons $A_i$ ($i =1,2,3$), $B$, and $Z'$ 
belonging to the gauge factors $SU(2)_L$, $U(1)_Y$, and $U(1)_N$ 
respectively is given by
\begin{eqnarray}
{\cal L} &=& |(\partial^\mu - {{i g_2} \over 2} \tau_i A_i^\mu + 
{{i g_1} \over 2} B^\mu + {{3 i g_N} \over {2 \sqrt {10}}} Z'^\mu) \tilde 
\Phi_1 |^2 \nonumber \\ &+& |(\partial^\mu - {{i g_2} \over 2} \tau_i A_i^\mu 
- {{i g_1} \over 2} B^\mu + {{i g_N} \over \sqrt {10}} Z'^\mu) \Phi_2 |^2 
\nonumber \\ &+& |(\partial^\mu - {{5 i g_N} \over {2 \sqrt {10}}} Z'\mu) 
\chi |^2,
\end{eqnarray}
where $\tau_i$ are the usual $2 \times 2$ Pauli matrices and the gauge 
coupling $g_N$ has been normalized to equal $g_2$ in the 
E$_6$ symmetry limit.  Let $\langle \phi_{1,2}^0 \rangle = v_{1,2}$ and 
$\langle \chi \rangle = u$, then for
\begin{equation}
W^\pm = {1 \over \sqrt 2} (A_1 \mp i A_2), ~~~ Z = {{g_2 A_3 - g_1 B} \over 
\sqrt {g_1^2 + g_2^2}},
\end{equation}
we have $M_W^2 = (1/2) g_2^2 (v_1^2 + v_2^2)$, and the mass-squared matrix 
spanning $Z$ and $Z'$ is given by
\begin{equation}
{\cal M}^2_{Z,Z'} = \left[ \begin{array} {c@{\quad}c} (1/2) g_Z^2 
(v_1^2 + v_2^2) & (g_N g_Z / 2 \sqrt {10}) ( -3 v_1^2 + 2 v_2^2) 
\\ (g_N g_Z/ 2 \sqrt {10}) ( -3 v_1^2 + 2 v_2^2) & (g_N^2/20) ( 
25 u^2 + 9 v_1^2 + 4 v_2^2) \end{array} 
\right],
\end{equation}
where $g_Z \equiv \sqrt {g_1^2 + g_2^2}$.

Let the mass eigenstates of the $Z-Z'$ system be
\begin{equation}
Z_1 = Z \cos \theta + Z' \sin \theta, ~~~ Z_2 = - Z \sin \theta + Z' \cos 
\theta,
\end{equation}
then the experimentally observed neutral gauge boson is identified in this 
model as $Z_1$, with mass given by
\begin{equation}
M^2_{Z_1} \equiv M^2_Z \simeq {1 \over 2} g_Z^2 v^2 \left[ 1 - \left( 
\sin^2 \beta - {3 \over 5} \right)^2 {v^2 \over u^2} \right],
\end{equation}
where
\begin{equation}
v^2 \equiv v_1^2 + v_2^2, ~~~ \tan \beta \equiv {v_2 \over v_1},
\end{equation}
and
\begin{equation}
\theta \simeq - \sqrt {2 \over 5} {g_Z \over g_N} \left( \sin^2 \beta - 
{3 \over 5} \right) {v^2 \over u^2}.
\end{equation}
The interaction Lagrangian of $Z_1$ with the leptons is now given by
\begin{eqnarray}
{\cal L} &=& \left( {1 \over 2} g_Z \cos \theta + {g_N \over \sqrt {10}} \sin 
\theta \right) \bar \nu_L \gamma_\mu \nu_L Z_1^\mu \nonumber \\ &+& \left( 
( -{1 \over 2} + \sin^2 \theta_W ) g_Z \cos \theta + {g_N \over \sqrt {10}} 
\sin \theta \right) \bar e_L \gamma_\mu e_L Z_1^\mu \nonumber \\ &+& \left( 
(\sin^2 \theta_W) g_Z \cos \theta - {g_N \over {2 \sqrt {10}}} \sin \theta 
\right) \bar e_R \gamma_\mu e_R Z_1^\mu,
\end{eqnarray}
where the subscripts $L(R)$ refer to left(right)-handed projections and 
$\sin^2 \theta_W = g_1^2 / g_Z^2$ is the usual electroweak mixing 
parameter of the standard model.  Using the leptonic widths and the 
forward-backward asymmetries, the deviations from the standard model are 
conveniently parametrized\cite{6}:
\begin{eqnarray}
\epsilon_1 &=& \left( \sin^4 \beta - {9 \over 25} \right) {v^2 \over u^2} ~=~ 
\alpha T, \\ \epsilon_2 &=& \left( \sin^2 \beta - {3 \over 5} \right) 
{v^2 \over u^2} ~=~ - {{\alpha U} \over {4 \sin^2 \theta_W}}, \\ \epsilon_3 
&=& {2 \over 5} \left( 1 + {1 \over {4 \sin^2 \theta_W}} \right) \left( 
\sin^2 \beta - {3 \over 5} \right) {v^2 \over u^2} ~=~ {{\alpha S} \over 
{4 \sin^2 \theta_W}},
\end{eqnarray}
where $\alpha$ is the electromagnetic fine-structure constant.  In the above 
we have also indicated how $Z-Z'$ mixing as measured in the lepton sector 
would affect the oblique $S,T,U$ parameters defined originally for the 
gauge-boson self energies only\cite{7}.  The present precision data from 
LEP at CERN are consistent with the standard model but the experimental 
error bars are of order a few $\times~10^{-3}$\cite{8}.  This 
means that $u \sim$ TeV is allowed.  Note also that the relative sign of 
$\epsilon_{1,2,3}$ is necessarily the same in this model.

\section{Two-Higgs-Doublet Sector}

The Higgs superfields of this model $(\nu_E, E)$, $(E^c, N_E^c)$, and $S$ 
are such that the term $f(\nu_E N_E^c - E E^c)S$ is the only allowed one 
in the superpotential.  This means that a supersymmetric mass term for $S$ 
is not possible and for $U(1)_N$ to be spontaneously broken, the 
supersymmetry must also be broken.  Consider now the Higgs potential.  The 
quartic terms are given by the sum of
\begin{equation}
V_F = |f|^2 [ (\Phi_1^\dagger \Phi_2)(\Phi_2^\dagger \Phi_1) + (\Phi_1^\dagger 
\Phi_1 + \Phi_2^\dagger \Phi_2)(\bar \chi \chi) ],
\end{equation}
and
\begin{eqnarray}
V_D &=& {1 \over 8} g_2^2 [ (\Phi_1^\dagger \Phi_1)^2 + (\Phi_2^\dagger 
\Phi_2)^2 + 2 (\Phi_1^\dagger \Phi_1)(\Phi_2^\dagger \Phi_2) - 4 
(\Phi_1^\dagger \Phi_2)(\Phi_2^\dagger \Phi_1) ] \nonumber \\ &+& 
{1 \over 8} g_1^2 [ (\Phi_1^\dagger \Phi_1)^2 + (\Phi_2^\dagger \Phi_2)^2 
- 2 (\Phi_1^\dagger \Phi_1)(\Phi_2^\dagger \Phi_2) ] \nonumber \\ &+& 
{1 \over 80} g_N^2 [ 9 (\Phi_1^\dagger \Phi_1)^2 + 4 (\Phi_2^\dagger \Phi_2)^2 
+ 12 (\Phi_1^\dagger \Phi_1)(\Phi_2^\dagger \Phi_2) - 30 (\Phi_1^\dagger 
\Phi_1)(\bar \chi \chi) \nonumber \\ &~& ~~~~~~~ - 20 (\Phi_2^\dagger \Phi_2)
(\bar \chi \chi) + 25 (\bar \chi \chi)^2 ].
\end{eqnarray}
The soft terms which also break the supersymmetry are given by
\begin{equation}
V_{soft} = \mu_1^2 \Phi_1^\dagger \Phi_1 + \mu_2^2 \Phi_2^\dagger \Phi_2 
+ m^2 \bar \chi \chi + fA \Phi_1^\dagger \Phi_2 \chi + (fA)^* \bar \chi 
\Phi_2^\dagger \Phi_1.
\end{equation}

The first stage of symmetry breaking occurs with $\langle \chi \rangle = u$. 
From $V_{soft}$ and $V_D$, we see that $u^2 = -8m^2/5g_N^2$.  Consequently, 
$\sqrt 2 Im \chi$ combines with $Z'$ to form a massive vector gauge boson 
and $\sqrt 2 Re \chi$ is a massive scalar boson.  Both have the same mass:
\begin{equation}
M_{Z'}^2 = m_\chi^2 = {5 \over 4} g_N^2 u^2.
\end{equation}
The reduced Higgs potential involving only the two doublets is then of the 
standard form:
\begin{eqnarray}
V &=& m_1^2 \Phi_1^\dagger \Phi_1 + m_2^2 \Phi_2^\dagger \Phi_2 + m_{12}^2 
(\Phi_1^\dagger \Phi_2 + \Phi_2^\dagger \Phi_1) \nonumber \\ &+& {1 \over 2} 
\lambda_1 (\Phi_1^\dagger \Phi_1)^2 + {1 \over 2} \lambda_2 (\Phi_2^\dagger 
\Phi_2)^2 + \lambda_3 (\Phi_1^\dagger \Phi_1)(\Phi_2^\dagger \Phi_2) + 
\lambda_4 (\Phi_1^\dagger \Phi_2)(\Phi_2^\dagger \Phi_1),
\end{eqnarray}
where
\begin{equation}
m_1^2 = \mu_1^2 - {3 \over 8} g_N^2 u^2, ~~~ m_2^2 = \mu_2^2 - {1 \over 4} 
g_N^2 u^2, ~~~ m_{12}^2 = fAu,
\end{equation}
assuming that $f$ and $A$ are real for simplicity.  In the above, we have 
of course also assumed implicitly that $m_1^2$, $m_2^2$, and $m_{12}^2$ are 
all small in magnitude relative to $u^2$.  The quartic scalar couplings 
$\lambda_{1,2,3,4}$ receive contributions not only from the coefficients of 
the corresponding terms in $V_D$ and $V_F$, but also from the cubic couplings 
of $\sqrt 2 Re \chi$ to the doublets which are proportional to $u$, as shown 
in Fig.~1.  As a result\cite{9},
\begin{eqnarray}
\lambda_1 &=& {1 \over 4} (g_1^2 + g_2^2) + {9 \over 40} g_N^2 - {{8 (f^2 - 
3 g_N^2/8)^2} \over {5 g_N^2}} ~=~ {1 \over 4} (g_1^2 + g_2^2) + {6 \over 5} 
f^2 - {{8 f^4} \over {5 g_N^2}}, \\ \lambda_2 &=& {1 \over 4} (g_1^2 + g_2^2) 
+ {1 \over 10} g_N^2 - {{8(f^2 - g_N^2/4)^2} \over {5 g_N^2}} ~=~ {1 \over 4} 
(g_1^2 + g_2^2) + {4 \over 5} f^2 - {{8 f^4} \over {5 g_N^2}}, \\ 
\lambda_3 &=& - {1 \over 4} g_1^2 + {1 \over 4} g_2^2 + {3 \over 20} g_N^2 - 
{{8(f^2 - 3g_N^2/8)(f^2 - g_N^2/4)} \over {5 g_N^2}} \nonumber \\ 
~&=& -{1 \over 4} g_1^2 + {1 \over 4} g_2^2 + f^2 - {{8 f^4} \over {5 g_N^2}}, 
\\ \lambda_4 &=& -{1 \over 2} g_2^2 + f^2.
\end{eqnarray}
It is obvious from the above that the two-Higgs-doublet sector of this model 
differs from that of the minimal supersymmetric standard model (MSSM) and 
reduces to the latter only in the limit $f=0$.  Note that if $m_{12}^2$ is 
of order $m_\chi^2$, then it is not consistent to assume that both $\Phi_1$ 
and $\Phi_2$ are light.  In that case, only a linear combination of $\Phi_1$ 
and $\Phi_2$ may be light and the electroweak Higgs sector reduces to that of 
just one doublet, as in the minimal standard model.

Since $V$ of Eq.~(22) should be bounded from below, we must have
\begin{equation}
\lambda_1 > 0, ~~~ \lambda_2 > 0, ~~~ \lambda_1 \lambda_2 - (\lambda_3 + 
\lambda_4)^2 > 0 ~~{\rm if}~ \lambda_3 + \lambda_4 < 0.
\end{equation}
Hence $f^2$ has an upper bound.  For $g_N^2 = (5/3) g_1^2$ which is a very 
good approximation if $U(1)_Y$ and $U(1)_N$ are unified only at a very high 
energy scale, we find that the ratio $f^2/g_Z^2$ has to be less than about  
0.35.  After electroweak symmetry breaking, the upper bound on the lighter 
of the two neutral scalar Higgs bosons is given in general by
\begin{equation}
(m_h^2)_{max} = 2 v^2 [\lambda_1 \cos^4 \beta + \lambda_2 \sin^4 \beta + 
2 (\lambda_3 + \lambda_4) \sin^2 \beta \cos^2 \beta] + \epsilon,
\end{equation}
where $\epsilon$ comes from radiative corrections, the largest contribution 
being that of the top quark:
\begin{equation}
\epsilon \simeq {{3 g_2^2 m_t^4} \over {8 \pi^2 M_W^2}} \ln \left( 1 + 
{\tilde m^2 \over m_t^2} \right),
\end{equation}
with $\tilde m$ coming from soft supersymmetry breaking.  In the present 
model, this becomes
\begin{equation}
(m_h^2)_{max} = 2 v^2 \left[ {1 \over 4} g_Z^2 \cos^2 2 \beta + f^2 \left( 
{3 \over 2} + {1 \over 5} \cos 2 \beta - {1 \over 2} \cos^2 2 \beta \right) 
- {{8 f^4} \over {5 g_N^2}} \right] + \epsilon.
\end{equation}
Considered as a function of $f^2$, the above quantity 
is maximized at
\begin{equation}
f_0^2 = {{5 g_N^2} \over 16} \left( {3 \over 2} + {1 \over 5} \cos 2 \beta 
- {1 \over 2} \cos^2 2 \beta \right).
\end{equation}
Assuming that $g_N^2 = (5/3) g_1^2$ as before, we find $f_0^2/g_Z^2$ 
to be always smaller than the upper bound 
we obtained earlier from requiring $V > 0$. 
Hence we plot $(m_h)_{max}$ in Fig.~2 for $f = f_0$ and $f=0$ as functions of 
$\cos^2 \beta$, as the maximum allowed values of $m_h$ in 
this model and in the MSSM respectively.  It is seen that for $m_t = 175$ 
GeV and $\tilde m = 1$ TeV, $m_h$ may be as high as 140 GeV in this model, 
as compared to 128 GeV in the MSSM.

For the charged Higgs boson $H^\pm = \sin \beta \phi_1^\pm - \cos \beta 
\phi_2^\pm$ and the pseudoscalar Higgs boson $A = \sqrt 2 (\sin \beta Im 
\phi_1^0 - \cos \beta Im \phi_2^0)$, we now have the sum rule
\begin{equation}
m^2_{H^\pm} = m_A^2 + M_W^2 - f^2 v^2,
\end{equation}
where $m_A^2 = - m^2_{12} / \sin \beta \cos \beta$.  Note that the above 
equation is common to all extensions\cite{9} of the MSSM with the term 
$f \Phi_1^\dagger \Phi_2 \chi$ in the superpotential and would serve as 
an unambiguous signal of physics beyond the MSSM at the supersymmetry 
breaking scale.

\section{The Neutralino Sector}

In the MSSM, there are four neutralinos (two gauge fermions and two Higgs 
fermions) which mix in a well-known $4 \times 4$ mass matrix\cite{10}.
Here we have six neutralinos: the gauginos of $U(1)_Y$ and the third 
component of $SU(2)_L$, the Higgsinos of $\bar \phi_1^0$ and $\phi_2^0$, 
the $U(1)_N$ gaugino and the $\chi$ Higgsino.  The corresponding mass 
matrix is then given by
\begin{equation}
{\cal M}_{\cal N} = \left[ \begin{array} {c@{\quad}c@{\quad}c@{\quad}c@
{\quad}c@{\quad}c} M_1 & 0 & -g_1 v_1 / \sqrt 2 & g_1 v_2 / \sqrt 2 & 0 & 0 \\ 
0 & M_2 & g_2 v_1 / \sqrt 2 & -g_2 v_2 / \sqrt 2 & 0 & 0 \\ -g_1 v_1 / \sqrt 2 
& g_2 v_1 / \sqrt 2 & 0 & f u & -3 g_N v_1/ 2 \sqrt 5 & f v_2 \\ g_1 v_2 / 
\sqrt 2 & -g_2 v_2 / \sqrt 2 & f u & 0 & -g_N v_2 / \sqrt 5 & f v_1 \\ 
0 & 0 & -3 g_N v_1 / 2 \sqrt 5 & -g_N v_2 / \sqrt 5 & M_1 & \sqrt 5 g_N u/2 \\ 
0 & 0 & f v_2 & f v_1 & \sqrt 5 g_N u/2 & 0 \end{array} \right],
\end{equation}
where $M_{1,2}$ are allowed $U(1)$ and $SU(2)$ gauge-invariant Majorana mass 
terms which break the supersymmetry softly.  Note that without the last two 
rows and columns, the above matrix does reduce to that of the MSSM if $fu$ 
is identified with $-\mu$.  Recall that if $f$ is very small, then the 
two-Higgs-doublet sector of this model is essentially indistinguishable 
from that of the MSSM, but now a difference will show up in the neutralino 
sector unless the $\mu$ parameter of the MSSM accidentally also happens to 
be very small.  In other words, there is an important correlation between 
the Higgs sector and the neutralino sector of this model which is not 
required in the MSSM.

Since $g_N u$ cannot be small, the neutralino mass matrix $\cal M_N$ 
reduces to either a $4 \times 4$ or $2 \times 2$ matrix, depending on 
whether $f u$ is small or not.  In the former case, it reduces to that of 
the MSSM but with the stipulation that the $\mu$ parameter must be small, 
{\it i.e.} of order 100 GeV.  This means that the two gauginos mix 
significantly with the two Higgsinos and the lightest supersymmetric 
particle (LSP) is likely to have nonnegligible components from all four 
states.  In the latter case, the effective $2 \times 2$ mass matrix becomes
\begin{equation}
{\cal M'_N} = \left[ \begin{array} {c@{\quad}c} M_1 + g_1^2 v_1 v_2 /f u & 
-g_1 g_2 v_1 v_2 /f u \\ -g_1 g_2 v_1 v_2 /f u & M_2 + g_2^2 v_1 v_2 /f u 
\end{array} \right].
\end{equation}
Since $v_1 v_2 / u$ is small, the mass eigenstates of $\cal M'_N$ are 
approximately the gauginos $\tilde B$ and $\tilde A_3$, with masses $M_1$ 
and $M_2$ respectively.  In supergravity models,
\begin{equation}
M_1 = {{5 g_1^2} \over {3 g_2^2}} M_2 \simeq 0.5~M_2,
\end{equation}
hence $\tilde B$ would be the LSP.

In the chargino sector, the corresponding mass matrix is
\begin{equation}
{\cal M_\chi} = \left[ \begin{array} {c@{\quad}c} M_2 & g_2 v_2 \\ g_2 v_1 & 
-f u \end{array} \right].
\end{equation}
If $f u$ is small, then both charginos can be of order 100 GeV, but if $f u$ 
is large (say of order 1 TeV), then only one may be light and its mass would 
be $M_2$.  In the MSSM, the superpotential has the allowed term 
$\mu \Phi_1^\dagger \Phi_2$.  
Hence there is no understanding as to why $\mu$ should be of order of 
the supersymmetry breaking scale, and not in fact very much greater.  Here 
$f u$ is naturally of order of the $U(1)_N$ breaking scale, and since the 
latter cannot be broken without also breaking supersymmetry, the two scales 
are necessarily equivalent.  This solves the so-called $\mu$ problem of the 
MSSM.

\section{Gauge-Coupling Unification}

In the MSSM, the three gauge couplings $g_3$, $g_2$, and $g_Y = 
(5/3)^{1 \over 2} g_1$ have been shown to converge to a single value at 
around $10^{16}$ GeV\cite{11}.  In the present model, with particle content 
belonging to complete {\bf 27} representations of E$_6$ and nothing else, this 
unification simply does not occur.  This is a general phenomenon of all 
grand unified models:  the experimental values of the three known gauge 
couplings at the electroweak energy scale are not compatible with a single 
value at some higher scale unless the particle content (excluding the gauge 
bosons) has different total contributions to the evolution of each coupling 
as a function of energy scale.  The evolution equations of $\alpha_i \equiv 
g_1^2 / 4 \pi$ are generically given to two-loop order by
\begin{equation}
\mu {{\partial \alpha_i} \over {\partial \mu}} = {1 \over {2 \pi}} \left[ 
b_i + {b_{ij} \over {4 \pi}} \alpha_j (\mu) \right] \alpha_i^2 (\mu),
\end{equation}
where $\mu$ is the running energy scale and the coefficients $b_i$ and 
$b_{ij}$ are determined by the particle content of the model.  To one loop, 
the above equation is easily solved:
\begin{equation}
\alpha_i^{-1} (M_1) = \alpha_i^{-1} (M_2) - {b_i \over {2 \pi}} \ln {M_1 
\over M_2}.
\end{equation}
Below $M_{\rm SUSY}$, assume the standard model with two Higgs doublets, then
\begin{equation}
b_Y = {21 \over 5}, ~~~ b_2 = -3, ~~~ b_3 = -7.
\end{equation}
Above $M_{\rm SUSY}$ in the MSSM,
\begin{equation}
b_Y = 3(2) + {3 \over 5} (4) \left( {1 \over 4} \right), ~~~ b_2 = -6 + 3(2) 
+ 2 \left( {1 \over 2} \right), ~~~ b_3 = -9 + 3(2).
\end{equation}
Note that in the above, the three supersymmetric 
families of quarks and leptons contribute equally to each coupling, 
whereas the two supersymmetric Higgs doublets do not.  The reason is 
that the former belong to complete representations of $SU(5)$ but not the 
latter.  For $M_{SUSY} \sim 10^4$ GeV, the gauge couplings would then unify 
at $M_U \sim 10^{16}$ GeV in the MSSM.

In the present model as it is, the one-loop coefficients of Eq.~(38) above 
$M_{\rm SUSY} (\sim u)$ are
\begin{equation}
b_Y = 3(3), ~~~ b_2 = -6 + 3(3), ~~~ b_3 = -9 + 3(3), ~~~ b_N = 3(3),
\end{equation}
because there are three complete {\bf 27} supermultiplets of E$_6$.  
[Actually $N$ is superheavy but it transforms trivially under 
$SU(3)_C \times SU(2)_L \times U(1)_Y \times U(1)_N$.]  To achieve 
gauge-coupling unification, we must add new particles in a judicious 
manner.  One possibility is to mimic the MSSM by adding one extra copy of 
the anomaly-free combination $(\nu_e, e)$ and $(E^c, N_E^c)$.  Then
\begin{equation}
\Delta b_Y = {3 \over 5}, ~~~ \Delta b_2 = 1, ~~~ \Delta b_3 = 0, ~~~ 
\Delta b_N = {2 \over 5}.
\end{equation}
Since the relative differences of $b_Y$, $b_2$, and $b_3$ are now the same as 
in the MSSM, we have again unification at $M_U \sim 10^{16}$ GeV, from which 
we can predict the value of $g_N$ at $M_{SUSY}$.  We show in Fig.~3 the 
evolution of $\alpha_i^{-1}$ using also the two-loop coefficients
\begin{equation}
b_{ij} = \left[ \begin{array} {c@{\quad}c@{\quad}c@{\quad}c} 
{234\over 25} & {54\over 5} & {84\over 5} & {339\over 100} \\ 
{18\over 5} & 39 & 24 & {73\over 20} \\
 3 & 9 & 48 & 3 \\
 {339\over 100} & 219\over 20 & 24 & {1897\over 200} 
\end{array} \right].
\end{equation}
We work in the ${\overline {\rm MS}}$ scheme, and take the two-loop matching
conditions accordingly\cite{wh}.   As an example, we use $\alpha =1/127.9$,
$\sin^2{\theta}_W =0.2317$, and
$\alpha_s =0.116$ at the scale $M_Z=91.187$ GeV.  We also choose $M_{SUSY}=1$
TeV and  use the top quark mass $m_t=175$ GeV. Note that the value of $\alpha_N$
is always close to that of $\alpha_Y$ since their one-loop beta fuctions are
close in value to each other and they are required to be unified at the scale
$M_U$.

Another possibility is to exploit the allowed variation of particle masses 
near the superstring scale of $M_U\approx 7 g_{U}\cdot 10^{17}\, {\rm
GeV} $ \cite{vsk} in the ${\overline {\rm MS}}$ scheme.  Just as Yukawa 
couplings are assumed to be subject only to the constraints of the unbroken 
gauge symmetry, the masses of the superheavy {\bf 27} and {\bf 27$^*$} 
multiplet components may also be allowed to vary accordingly.  For example, 
take three copies of $(u,d) + (u^*,d^*)$ and
$(\nu_e,e) + (\nu_e^*,e^*)$  with $M'$ much below $M_U$, then between $M'$ and
$M_U$,
\begin{equation}
\Delta b_Y = 3 \times \left( {1 \over 5} + {3 \over 5} \right) = {12 \over 5}, 
~~~ \Delta b_2 = 3 \times (3 + 1) = 12,
\end{equation}
\begin{equation}
\Delta b_3 = 3 \times (2 + 0) = 6, ~~~ \Delta b_N = 3 \times \left( 
{3 \over 10} + {2 \over 5} \right) = {21 \over 10}.
\end{equation}
For $M' \sim 10^{16}$ GeV, gauge-coupling unification at $M_U \sim 7 \times 
10^{17}$ GeV is again achieved.  We show in Fig.~4 the evolution of 
$\alpha_i^{-1}$ using also the two-loop coefficients
\begin{equation}
b_{ij} = \left[ \begin{array} {c@{\quad}c@{\quad}c@{\quad}c} 
9 & 9 & {84\over 5} & 3 \\ 
3 & 39 & 24 & 3 \\
3 & 9 & 48 & 3 \\ 
3 & 9 & 24 & 9 
\end{array} \right]
\end{equation}
between $M_{\rm SUSY}$ and $M'$, and 
\begin{equation}
b_{ij} = \left[ \begin{array} {c@{\quad}c@{\quad}c@{\quad}c} 
{253\over 25} & {81\over 5} & 20 & {102\over 25} \\ 
{26\over 5} & 123 & 72 & {51\over 100} \\ 
3 & 27 & 116 & {18\over 5} \\ 
{102\over 25} & {153\over 10} & 24 & {957\over 100} 
\end{array} \right]
\end{equation}
between $M'$ and $M_U$. As an example,  we use $\alpha =1/127.9$,
$\sin^2{\theta}_W =0.2317$, and $\alpha_s =0.123\pm 0.006$ at the scale
$M_Z=91.187$ GeV, $M_{\rm SUSY}=1$ TeV and the top quark mass $m_t=175$ GeV. For
$\alpha_s (M_Z) =0.123$, we find 
$M'=5.9\times 10^{16}$ GeV.  It should be emphasized that the sharp turn at 
$M'$ should not be taken too literally but only as an indication that 
gauge couplings may in fact evolve drastically near the unification energy 
scale.  This possibility allows us to have unification without having split 
multiplets containing both superheavy and light components, as in most 
grand unified models. The size of $\alpha_N$ is always very close to that of
$\alpha_Y$ since they have the same one-loop beta functions for scales beneath
$M'$ and are required to be unified at $M_U$. We note that the two-loop
corrections are larger here than in the MSSM due to the much larger particle
content. We also observe that for
$\alpha_s (M_Z) =0.123$ we obtain an
$M_U$ which is about 1.5 times the  superstring scale of $7 g_{U}\cdot
10^{17}\,{\rm GeV}$, whereas the $M_U$  in most supersymmetric grand unified
models is about 0.04 times that number. 

\section{Concluding Remarks}

To accommodate a naturally light singlet neutrino, an extra U(1) factor is 
called for.  It has been shown\cite{4} that the superstring-inspired E$_6$ 
model is tailor-made for this purpose as it contains $U(1)_N$ which has 
exactly the required properties.  To obtain $U(1)_N$ as an unbroken gauge 
group, we need to break E$_6$ spontaneously along the $N$ and $N^*$ 
directions with superheavy {\bf 27}'s and {\bf 27$^*$}'s while preserving 
supersymmetry.  This is impossible if the superpotential is allowed only terms 
up to cubic order so that the theory is renormalizable.  On the other hand, 
the requirement of renormalizability may not be applicable at the superstring 
unification scale, in which case the quartic term $M^{-1}$ 
{\bf 27 27$^*$ 27 27$^*$} in conjunction with the quadratic term $m$ 
{\bf 27 27$^*$} in the superpotential would result in $\langle$ {\bf 27} 
$\rangle$ = $\langle$ {\bf 27$^*$} $\rangle$ = $(-2mM)^{1 \over 2}$ without 
breaking supersymmetry.

The addition of $U(1)_N$ has several other interesting phenomenological 
consequences.  (1) The $U(1)_N$ neutral gauge boson Z' mixes with the 
standard-model Z and affects the precision data at LEP.  From the present 
experimental error bars on the $\epsilon_{1,2,3}$ parameters, we find that the 
$U(1)_N$ breaking scale could be as low as a few TeV.  (2) The spontaneous 
breaking of $U(1)_N$ is accomplished only with the presence of a mass term 
in the Higgs potential which breaks the supersymmetry softly.  Hence the 
reduced two-doublet Higgs potential at the electroweak energy scale is not 
guaranteed to be that of the MSSM.  In fact, the scalar quartic couplings now 
depend also on a new Yukawa coupling $f$ as well as the gauge coupling $g_N$. 
Assuming that $g_N = (5/3)^{1 \over 2} g_1$, one result is that the upper 
bound on the lighter of the two neutral scalar Higgs bosons is now 140 GeV 
instead of 128 GeV in the MSSM.  (3) The neutralino mass matrix also 
depends on $f$, hence there is a correlation here with the Higgs sector.  
Such a connection is not present in the MSSM.  (4) This model 
may also be compatible with gauge-coupling unification.  We identify two 
possible scenarios.  One is just like the MSSM with two light doublets 
presumably belonging to complete multiplets (of the grand unified group) 
whose other members are superheavy; the other requires no light-heavy 
splitting but assumes a large variation of superheavy masses near the 
unification scale.
\vspace{0.3in}
\begin{center} {ACKNOWLEDGEMENT}
\end{center}

This work was supported in part by the U.~S.~Department of Energy under 
Grant No.~DE-FG03-94ER40837.

\newpage
\bibliographystyle{unsrt}

\newpage
\leftline{{\Large\bf Figure Captions}}
\begin{itemize}

\item[Fig. 1~:]
{{The tree-level Feynman diagrams due to the cubic couplings of the scalar 
$\sqrt{2} {\rm Re}\chi$ to the scalar Higgs doublets $\Phi_{1,2}$  which 
contribute to the quartic scalar couplings
$\lambda_{1,2,3,4}$ of the reduced Higgs potential given in Eq.
(22).}\label{fig1}}

\item[Fig. 2~:] {{The upper bound on the mass of the lighter of the two 
neutral scalar Higgs bosons 
$\left( m_h\right)_{max}$ for $f=f_0$ and $f=0$ as functions of
$\cos^2{\beta}$, as the maximum allowed values of $m_h$ in the
model discussed here and in the MSSM respectively. We have  used $\alpha
=1/127.9$ and $\sin^2{\theta}_W =0.2317$ at the $M_Z$ scale,  $m_t = 175$ 
GeV, and $\tilde m = 1$ TeV.}\label{fig2}}

\item[Fig. 3~:] {{The two-loop evolution of the gauge couplings of
the unification  scenario involving three complete ${\bf 27}$ supermultiplets
and  one extra copy of $(\nu_e ,e)$ and $(E^c,N_E^c)$ with mass of order 
$M_{SUSY}$ as explained in the text. We have used $\alpha =1/127.9$, 
$\sin^2{\theta}_W =0.2317$, and
$\alpha_s =0.116$ at the scale
$M_Z=91.187$ GeV,  $M_{SUSY}=1$ TeV, and 
$m_t=175$ GeV.}\label{fig3}}

\item[Fig. 4~:] {{The two-loop evolution of the gauge couplings of
the unification scenario explained in the text which involves three complete
${\bf 27}$ supermultiplets for scales above
$M_{\rm SUSY}$ and  with three additional copies of $(u,d)+(u^*,d^*)$ and
$(\nu_e ,e)+(\nu_e^* ,e^*)$  with mass of order the intermediate scale $M'$. We
have used
$\alpha =1/127.9$,
$\sin^2{\theta}_W =0.2317$, and
$\alpha_s =0.123\pm 0.006$ at the scale
$M_Z=91.187$ GeV,  $M_{SUSY}=1$ TeV, and 
$m_t=175$ GeV. The dashed lines correspond to $\alpha_s 
(M_Z)=0.117$ and 0.129.}\label{fig4}}

\end{itemize}

\vfill

\end{document}